% Authors:   R. F. Bikbaev and R. A. Sharipov
% Comment:   AmSTeX, Ver. 2.1h, 5 pages, amsppt style,
% Published: Physics Letters A, 134 (1988), no. 2, 105-107.
%            Received 1 August 1988, accepted for publication
%            13 October 1988. Communicated by A. P. Fordy.
% Web-sites: http://www.geocities.com/CapeCanaveral/Lab/5341
%            http://www.bashedu.ru/sharipov
%
\input amstex
\documentstyle{amsppt}
\nopagenumbers
\def\diag{\operatorname{diag}}
\def\Jac{\operatorname{Jac}}
\def\divsr{\operatorname{div}}
\def\Real{\operatorname{Re}}
\pagewidth{360.0pt}
\pageheight{606.0pt}
\leftheadtext{R\.~F\.~Bikbaev and R\.~A\.~Sharipov}
\topmatter
\title MAGNETIZATION WAVES IN THE LANDAU-LIFSHITZ MODEL
\endtitle
\author
R\.~F\.~Bikbaev and R\.~A\.~Sharipov\footnotemark
\endauthor
\abstract
The solutions of the Landau-Lifshitz equation with finite-gap
behavior at infinity are considered. By means of the inverse
scattering method the large-time asymptotics is obtained.
\endabstract
\address
Mathematical Institute of Bashkir Scientific Center, Academy of
Sciences of the USSR, Tukaeva 50, 450057 Ufa, USSR
\endaddress
\endtopmatter
\loadbold
\document
\footnotetext{\vtop{\hsize 30em 
\leftline{http://www.geocities.com/CapeCanaveral/Lab/5341}
\leftline{http://www.bashedu.ru/sharipov}}}
     {\bf 1.} The Landau-Lifshitz equation \cite{1} describing 
the dynamics of the magnetization vector $\bold S$ for the 
one-dimensional ferromagnet of the ``light-plane'' type can be
written in the following form:
$$
\bold S_t=\bold S\times\bold S_{xx}+\bold S\times J\bold S,
\quad |\bold S|=1,\quad J=\diag(0,0,- 16\omega^2).
\tag{1}
$$
In \cite{2,\,3} equation \thetag{1} was shown to be completely 
integrable and it was represented as a compatibility condition for
the pair of linear equations
$$
\xalignat 2
&\partial_x\varPsi=U\varPsi,&&\partial_t\varPsi=V\varPsi
\tag{2}
\endxalignat
$$
with $2\times 2$ matrices $U$ and $V$ of the form
$$
\align
&U=-i\sum^3_{\alpha=1}S^\alpha\,w_\alpha\,\sigma_\alpha,\\
&V=2i\sum^3_{\alpha=1}\frac{w_1\,w_2\,w_3}{w_\alpha}\,
S^\alpha\,\sigma_\alpha-i\sum^3_{\alpha=1}[\bold S\times
\bold S_x]^\alpha\,w_\alpha\,\sigma_\alpha,
\endalign
$$
where $\sigma_\alpha$ are the Pauli matrices and $w_1=w_2=
\sqrt{\lambda^2-\omega^2}$, $w_3=\lambda$. Soliton-like solutions
of \thetag{1} are well-known (see \cite{4,\,5}). The class of
periodic and almost periodic wave-like solutions of \thetag{1}
contains an important subclass of algebro-geometric (or finite-gap)
solutions. They were constructed in \cite{5,\,6}. The study of
algebro-geometric solutions for integrable equations was initiated
by Novikov in \cite{7}, it led to the well-developed theory of
finite-gap integration (see review \cite{8}).\par
     In this paper we study the large-time asymptotics for ``nearly
finite-gap solutions'' of the Landau-Lifshitz equation, i\.~e\. the
solutions $\bold S$ with the following behavior as $x\to\pm\infty$:
$$
\aligned
&\bold S(x,t)\to S(x,t\,|\,\Gamma,D_1,\delta_1),\quad x\to+\infty,\\
&\bold S(x,t)\to S(x,t\,|\,\Gamma,D_2,\delta_2),\quad x\to-\infty.
\endaligned
\tag{3}
$$
Here $S(x,t\,|\,\Gamma,\delta)$ denotes a real smooth $g$-gap solution
of \thetag{1} with a phase $\delta$ constructed on a base of the
hyperelliptic Riemann surface $\Gamma$ with a fixed divisor
$D=P_1+\ldots+P_g$ on it. Given the branching points
$$
\lambda_0=-\omega<\lambda_1<\lambda_2<\ldots<\lambda_{2g}<\omega
=\lambda_{2g+1}
$$
of $\Gamma$ one can define the meromorphic function
$$
Y=\sqrt{(\lambda^2-\omega^2)(\lambda-\lambda_1)\cdot\ldots\cdot
(\lambda-\lambda_{2g})}
$$
on $\Gamma$ and the pair of infinity points $P^\pm_\infty$, with
$Y\sim\pm\lambda^{g+1}$ as $P\to P^\pm_\infty$. The Riemann surface
$\Gamma$ consists of two sheets: $\Gamma_+$ (upper sheet) and
$\Gamma_-$ (lower sheet). It admits of the hyperelliptic involution
$\sigma$, which does interchange sheets, and the antiholomorphic
involution $\tau$, $(\lambda(\tau P)=\overline{\lambda(P)},\quad
Y(\tau P)=-\overline{Y(P)})$, which does not. The boundary $\partial
\Gamma_+$ is a collection of $g$ cycles $\gamma_1,\,\ldots,\,\gamma_g$
and the cycle $\gamma_\infty$ passing through two infinity points 
$P^\pm_\infty$.\par
\vskip 18ex \hskip 10em See in separate file: {\bf Bikb.gif}.\vskip 18ex
%\vskip 2ex \ \ \ \ \special{em:graph Bikb.bmp}\vskip 34ex
% You can convert <.gif> to <.bmp> and uncomment upper commented line.
% In this case Resolution 300dpi and magnification 1000 recommended.
\centerline{Fig\.~1.}
\vskip 5ex
     Let us choose the canonical basis of cycles $a_i$, $b_i$, 
$i=l,\,\ldots,\,g$ on $\Gamma$ as it is shown on fig\.~1. The
finite-gap solution $S(x,t\,|\,\Gamma,\delta)$ then is given up to a
phase shift by explicit formulae in terms of Riemann
$\theta$-functions:
$$
\xalignat 3
&S^1=\frac{C_1C_2-C_3C_4}{C_3C_2-C_1C_4},
&&S^2=-i\frac{C_1C_2+C_3C_4}{C_3C_2-C_1C_4},
&&S^3=\frac{C_3C_2+C_4C_1}{C_3C_2-C_1C_4},
\endxalignat
$$
Here
$$
\xalignat 2
&C_1=\theta[n,0](\Omega+\Delta+z),&&C_3=\theta(\Omega+\Delta+z),\\
\vspace{1ex}
&C_2=-\theta[n,0](\Omega+\Delta-z),&&C_4=\theta(\Omega+\Delta-z),\\
&n=\frac{1}{2}(1,0,\ldots,0).
\endxalignat
$$
The change of phase $\delta$ \pagebreak is equivalent to the rotation 
of the vector $\bold S$ around the third coordinate axis. The vector 
$\Delta$ is connected with the divisor by the Abel map
$$
A\!:\divsr(\Gamma)\to\Jac(\Gamma),\quad
A_i(P)=\int\limits^{\,\,P}_{\lambda_0}\omega_i,\quad P\in\Gamma,
$$
according to the formula $\Delta=-A(D)-K$, where $K$ is the vector of
Riemann constants. Real solutions $\bold S(x,t)$ corresponding to real
divisors are determined by the restrictions
$$
A(D)-A(\tau D)=A(\lambda_0+\lambda_{2g+1}-P^+_\infty-P^-_\infty)=0.
\tag{4}
$$
Vector $Q=i(V^{(1)}x-V^{(2)}t)$ is composed of two vectors $V^{(1)}$
and $V^{(2)}$, being the vectors $b$-periods of two normalized abelian
differentials of the second kind with the only poles at infinities
$P^\pm_\infty$. These differentials have the following leading terms
of Laurent expansions at these points:
$$
\xalignat 2
&\Omega^{(1)}=\mp d\lambda+\ldots,
&&\Omega^{(2)}=\pm 4\,\lambda\,d\lambda+\ldots.
\endxalignat
$$
Vector $z\in\Jac(\Gamma)$ is equal to $A(P^+_\infty)$, the path 
of integration $\gamma$ is shown on fig\.~1.\par
     The reality condition \thetag{4} defines $2^g$ disjoint real tori
$T_\nu,\ \nu=0,\ldots,2^g-1$ in $\Jac(\Gamma)$. We choose only one of
them: torus $T_0$ with
$$
\Real[\Delta+A(\lambda_0)]=0,
$$
on which the $\theta$-function $\theta(A(\lambda_0)+\Omega+\Delta)$
does not vanish (see \cite{9}). The main instrument in constructing 
finite-gap solutions is the matrix Baker-Akhiezer function
$$
e(P)=\Vmatrix e^+_1(P) &e^+_1(\sigma P)\\
\vspace{3ex}
e^+_2(P) &e^+_2(\sigma P)\endVmatrix
$$
solving equations \thetag{2}. The first column of it is given up to a
scalar multiples $f_1(x,t)$ and $f_2(x,t)$ by formulas
$$
\aligned
&e^+_1(P)=f_1\,e^{i\delta/2}\,\frac{\theta(A(\lambda)+\Omega+\Delta)}
{\theta(A(\lambda)+\Delta)}\,\exp\left(i\int\limits^P_{\lambda_0}
(\Omega^{(1)}x+\Omega^{(2)}t)\right),\\
\vspace{2ex}
&e^+_1(P)=f_2\,e^{-i\delta/2}\,\frac{\theta[n,0](A(\lambda)+\Omega+\Delta)}
{\theta(A(\lambda)+\Delta)}\,\exp\left(i\int\limits^P_{\lambda_0}
(\Omega^{(1)}x+\Omega^{(2)}t)\right).
\endaligned
\tag{6}
$$
Multiples $f_1(x,t)$ and $f_2(x,t)$ are defined by fixing $\det e(P)$ and
by the condition $e_1(\lambda_0)/e_2(\lambda_{2g+1})=e^{i\delta}$.\par
     {\it Remark.} The torus $T_0$ is an exceptional real torus
in the following sense: Baker-Akhiezer function $e(P,x,t)$ is non-singular
bounded function in $x,t$ \pagebreak for $P\in\partial\Gamma_+$.\medskip
     {\bf 2.} In order to construct a scattering theory for $\bold
S(x,t)$ of the form \thetag{3} let us define the vectorial Jost functions
$\varPhi(P)$ and $\varPsi(P)$ solving \thetag{2} and having asymptotics
$$
\align
&\varPhi(P)\to e^+(P,D_1,\delta_1)\text{\ \ as \ }x\to +\infty,\\
&\varPsi(P)\to e^+(P,D_2,\delta_2)\text{\ \ as \ }x\to -\infty.
\endalign
$$
The functions $\varPhi$, $\varPsi$ are bounded with each
other by scattering data $a(P)$, $b(P)$:
$$
\varPhi(P)=\varPsi(P)\,a(P)+\varPsi(\sigma P)b(P),\quad
P\in\partial\Gamma_+.
\tag{7}
$$
In this paper we study the non-soliton case, i\.~e\. $a(P)\neq 0$,
if $P\in\Gamma_+$. Starting from \thetag{7} we obtain a scattering
theory for \thetag{1}, \thetag{3} most similar to that of \cite{10}
for the fast-decreasing case. The only difference consists in the
existence of relations between asymptotic divisors $D_1$, $D_2$,
phases $\delta_1$, $\delta_2$ and scattering data $a(P)$, $b(P)$:
$$
\aligned
&A(D_2-D_1)=\frac{1}{2\pi i}\int\limits_{\partial\Gamma_+}
\ln|1-r(P)\,r(\sigma P)|\,\omega(P),\\
\vspace{1ex}
&\delta_1-\delta_2=-i\,\ln\left(\frac{a(\lambda_{2g+1})+b(\lambda_{2g+1})}
{a(\lambda_0)-b(\lambda_0)}\right).
\endaligned
\tag{8}
$$
Here $r(P)=b(P)/a(P)$ is the reflection coefficient. It should be pointed
out that for our choice of divisors $D_1$ and $D_2$ (i\.\,e\. torus $T_0$)
$1-r(P)\,r(\sigma P)$ is a real and positive function on $\partial
\Gamma_+$.\par
     For the asymptotical analysis of \thetag{1}, \thetag{3} we use 
a singular integral equation for Jost functions similar to that of
\cite{11}. Our method is a generalization of the asymptotical
construction of \cite{12}.\par
     The final result of our investigation is the following: the main
term of the asymptotics for $S(x,t)$ as $t\to+\infty$ is given by the
finite-gap solution
$$
\bold S(x,t)=\bold S(x,t\,|\,D(\xi),\delta(\xi))+\varepsilon(\xi,t),
\quad\varepsilon(\xi,t)=o(1),
$$
with the phase $\delta(\xi)$ and divisor $D(\xi)$ depending on
the ``slow variable'' $\xi=x/t$ according to
$$
\aligned
&A(D(\xi))=A(D_2)-\frac{1}{2\pi i}\int\limits_{\ell(\xi)}
\ln|1-r(P)\,r(\sigma P)|\,\omega(P),\\
\vspace{1ex}
&\delta(\xi)=\delta_2-i\,\ln\left(\frac{\tilde A(\lambda_{2g+1})}
{\tilde A(\lambda_0)}\,\frac{1+\tilde r(\lambda_{2g+1})}
{1-\tilde r(\lambda_0)}\right).
\endaligned
\tag{10}
$$
Here the path of integration $\ell(\xi)$ is a part of the contour
$\partial\Gamma_+$ which is situated to the left of the unique
stationary point $P_0(\xi)$ (see fig\.~1) defined by the condition
$$
\left(\Omega^{(1)}\xi+\Omega^{(2)}\right)
\hbox{\vrule height 10pt depth 10pt width 0.5pt}_{\,P=P_0}=0.
$$
The function $\tilde A(P)$ is given by
$$
\align
&\qquad\qquad\tilde A(P)=\lim\limits_{P'\to P}\alpha(P'),
\quad P'\in\Gamma_+,\quad P\in\partial\Gamma_+,\\
\vspace{1ex}
&\alpha(P)=\frac{\theta((A(P)-A(D(\xi))-K)}{\theta((A(P)-A(D_2)-K)}\,
\exp\left(-\frac{1}{2\pi i}\shave{\int\limits_{\ell(\xi)}}
M(Q,P)\,\ln(1-r(Q)\,r(\sigma Q))\right),
\endalign
$$
where $M(Q,P)$ is the multivalued Cauchy kernel (see \cite{9}). The 
function $\tilde r(P)$ is given by
$$
\tilde r(P)=\cases r(P) &\text{for \ }P\in\ell(\xi),\\
0 &\text{for \ }P\notin\ell(\xi).\endcases
$$
The value of the rest term of asymptotics \thetag{9} depends on
whether $P_0\in\partial\Gamma_+$ or not. In the first case
$\varepsilon=O(t^{-1/2})$, in the second case $\varepsilon=
o(t^{-N})$ for any $N>0$. The scattering problem studied here
describes the interaction of two magnetization waves with the
same spectrum $\Gamma$. After finishing all the ``transition
processes'' two interacting waves consolidate into one asymptotical
wave \thetag{9} with slowly changing phases.

\enddocument
\end